\documentclass[final,5p,times,twocolumn,authoryear]{elsarticle} 

\usepackage{amssymb}

\usepackage{lineno}

\journal{Applied Radiation and Isotopes}

\begin{document}

\begin{frontmatter}



\title{Shielding concepts for low-background proportional counter arrays in surface laboratories}


\author{C.E.~Aalseth}
\author{P.H.~Humble}
\author{E.K.~Mace}
\author{J.L.~Orrell\corref{cor}}\ead{john.orrell@pnnl.gov}
\author{A.~Seifert}
\author{R.M.~Williams}

\cortext[cor]{Corresponding author.}

\address{Pacific Northwest Laboratory, Richland, WA 99352, USA}

\begin{abstract}
Development of ultra low background gas proportional counters has made the contribution from naturally occurring radioactive isotopes -- primarily $\alpha$ and $\beta$ activity in the uranium and thorium decay chains -- inconsequential to instrumental sensitivity levels when measurements are performed in above ground surface laboratories. Simple lead shielding is enough to mitigate against gamma rays as gas proportional counters are already relatively insensitive to naturally occurring gamma radiation. The dominant background in these surface laboratory measurements using ultra low background gas proportional counters is due to cosmic ray generated muons, neutrons, and protons. Studies of measurements with ultra low background gas proportional counters in surface and underground laboratories as well as radiation transport Monte Carlo simulations suggest a preferred conceptual design to achieve the highest possible sensitivity from an array of low background gas proportional counters when operated in a surface laboratory. The basis for a low background gas proportional counter array and the preferred shielding configuration is reported, especially in relation to measurements of radioactive gases having low energy decays such as $^{37}$Ar.
\end{abstract}


\end{frontmatter}



\section{Introduction}
\label{sec:Introduction}

Low background radiometric measurement systems are typically located in underground laboratories to shield against cosmic-ray products (e.g., protons, neutrons, and muons) that create background event rates and ultimately limit a system's sensitivity reach. The effort to further improve the sensitivity of such underground systems, in one case, has led to the development of ultra low background gas proportional counters made from ultra-pure electroformed copper. These underground proportional counters are further operated in shields to exclude naturally occurring radiation in the laboratory from contributing signals to the detector system. As a consequence of these development efforts the proportional counters now produced have a contribution from naturally occurring radioactive isotopes -- primarily $\alpha$ and $\beta$ activity in the uranium and thorium decay chains -- that is entirely inconsequential to instrumental sensitivity levels when the measurements performed in \emph{above-ground} surface laboratories. This allows for a single-minded focus on re-addressing the challenge of the cosmic-ray product induced background contribution in an above-ground proportional counter system. This article presents the rational and conceptual design for a proportional counter array designed for above-ground operation, based upon an assessment of the background contributions from both cosmic-ray products and naturally occurring radioactivity. As a concrete example, measurement of environmental naturally occurring levels of $^{37}$Ar is chosen as an application to explore.

\section{Proportional counter backgrounds}
\label{sec:ProportionalCounterBackgrounds}

Recent reported measurements and simulations presented below provide the basis for bracketing the expected performance of a ultra-low-background counter in a surface laboratory. This section describes the prior measurements and presents simulations used as a basis for inferring the detection system's expected sensitivity. To provide a concrete quantitative estimate, $^{37}$Ar detection sensitivity is evaluated. The 2.8~keV emission after $^{37}$Ar decay (\cite{Barsanov2007}) provides an uncomplicated nearly single-point energy deposition within the proportional count gas (\cite{Xiang2007}). Thus understanding the performance of the radiation detection is particularly simple allowing a straight forward discussion of background contributions. Furthermore the discussion of $^{37}$Ar measurements is motivated by a desire to understand the ultimate reach of surface laboratory systems (\cite{Xiang2008}, \cite{Xiang2009}) that ideally would have sensitivity to the naturally occurring levels of in the environment (\cite{Cao2009}, \cite{Riedmann2011}).

\subsection{Surface cosmic-ray background}
\label{subsubsec:SurfaceBackground}
\begin{figure}
\begin{center}
\includegraphics[width=0.91\columnwidth]{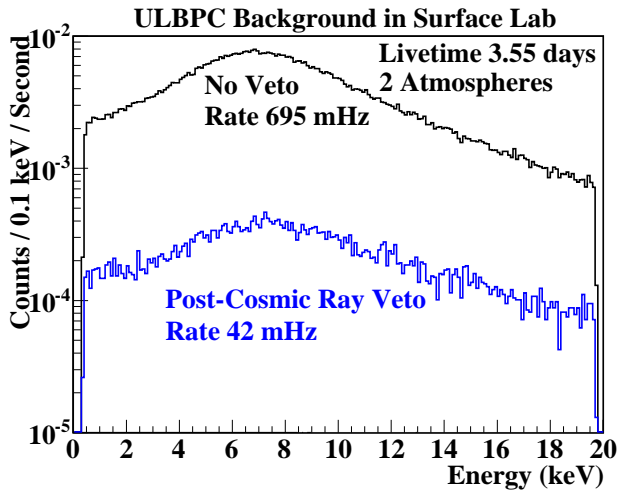}
\caption{\label{fig:ULBPCCosmicRayBackground} The background from cosmic rays passing through an ULBPC, before and after application of the PVT-based cosmic-ray veto. Data was collected in a small shield located in a surface laboratory.}
\end{center}
\end{figure}

A previous gas proportional measurement of $^{37}$Ar in a surface laboratory (\cite{Aalseth2011}) provides a point of reference for understanding the backgrounds of a single detector, surface laboratory gas proportional counter system. The single proportional counter used in that measurement was an ultra-low background proportional counter (ULBPC) fabricated from ultra-pure electroformed copper (\cite{Aalseth2009}). The ULBPC is roughly 9$^{\prime\prime}$ in length and 1$^{\prime\prime}$ in diameter with an internal volume of $\sim$0.1 L and was operated in a lead shield consisting of 2$^{\prime\prime}$, 4$^{\prime\prime}$, and 8$^{\prime\prime}$ thick walls (top, sides, and bottom, respectively). Two 2$^{\prime\prime}$ thick polyvinyl toluene (PVT) cosmic-ray veto panels were place above and below the shield, having 33$^{\prime\prime}$$\times$14$^{\prime\prime}$ and 56$^{\prime\prime}$$\times$19$^{\prime\prime}$ footprints, respectively. This shielding arrangement is \emph{not} ideal due to the absence of cosmic-ray veto panel coverage on four sides. Nevertheless, as reported (\cite{Aalseth2011}), the cosmic veto shield was able to reduce the background in the 1-20~keV measurement region by a factor of $\sim$16.5. These results are reproduced in Figure \ref{fig:ULBPCCosmicRayBackground} and show that while the cosmic-ray veto panels reduced the background from cosmic rays, the fact that the \emph{shape} of the background spectrum is unchanged implies the dominant background remains due to untagged (un-vetoed) cosmic ray products.

\subsection{Shallow underground cosmic-ray background}
\label{subsubsec:ShallowBackground}
Pacific Northwest National Laboratory (PNNL) operates a shallow underground laboratory in Richland, WA having an overburden of $\sim$30 meters water equivalent overburden shielding against cosmic rays (\cite{Aalseth2012}). The overburden of the underground laboratory acts as a shield against cosmic-ray products, stopping protons and reducing neutrons, both produced in the atmosphere by spallation processes. The cosmic-ray muon flux -- a more penetrating cosmic-ray product -- is reduced by $\sim$85\% in the underground laboratory and is used as the final quantitative metric for evaluating the effective depth of the underground laboratory (\cite{Aalseth2012}).

The relevance of the underground laboratory to a surface laboratory measurement system is as a point of comparison for an ideal measurement sensitivity level. Detectors of the same ULBPC design described in the previous section \ref{subsubsec:SurfaceBackground} were deployed in the shallow underground laboratory. As noted in the previous section, the dominant background for these proportional counters systems, when operated on the surface, is the residual cosmic-ray (direct or induced) backgrounds despite the use of scintillator-based cosmic-ray veto detectors. Thus the measurement sensitivity of a ULBPC measurement made in the shallow underground facility sets a practical lower bound on the potential sensitivity one could hope to achieve in a surface laboratory measurement system.

\begin{figure}
\begin{center}
\includegraphics[width=\columnwidth]{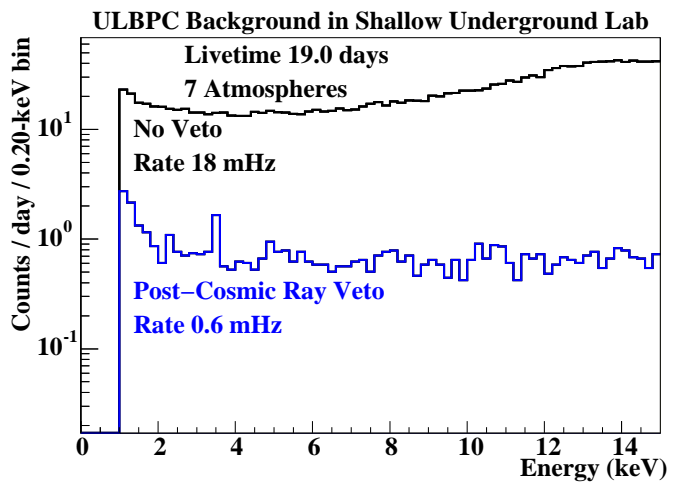}
\caption{\label{fig:ULBPCCosmicRayBackgroundShallowLab} The background in an ULBPC, before and after application of the scintillating plastic cosmic-ray veto. Data was collected in a proportional counter shield located in the PNNL shallow underground laboratory.}
\end{center}
\end{figure}

Figure \ref{fig:ULBPCCosmicRayBackgroundShallowLab} presents results from 19.0 live-days of background data collected from an ULBPC operated in the shallow underground laboratory (\cite{Aalseth2013}). The proportional counter was located within a low background shield consisting of: (1) a gas-tight enclosure, (2) an active anti-cosmic veto detector consisting of 5-cm-thick scintillating plastic panels covering five sides of the shield, (3) a layer of 30\% borated polyethylene with a thickness (2.5 cm) selected to provide opacity for thermal neutrons, (4) a passive lead layer (15 cm) to limit external gamma-ray induced interactions, and (5) a passive high-purity copper inner layer with a thickness (7.6 cm) to limit induced interactions from bremsstrahlung gamma-ray production due to $^{210}$Pb ($\sim$30~Bq/kg) in the passive lead layer. The proportional counter referred to here was operated at 7 atmospheres of pressure of P-10 counting gas.

The upper histogram of Fig. \ref{fig:ULBPCCosmicRayBackgroundShallowLab} is the gross detector counts while the lower histogram is the data remaining after the cosmic-ray veto system is employed to reject events (``vetoed data''). The post-cosmic ray veto data appears as a continuum background which is believed to be due to internal backgrounds (e.g. $\beta$, $\alpha$, and potentially neutrons) and the P-10 counting gas containing $^{39}$Ar ($\beta$-emitter: $E_{\beta}^{\mathrm{end}}$ $=$ 565~keV; $t_{1/2}$ $=$ 269~years). For comparison to the $^{37}$Ar sensitivity level achieved in the previous above group measurement ($\sim$45~mBq/SCM), the shallow underground laboratory ULBPC $^{37}$Ar measurement has an initial reported sensitivity of $\sim$2~mBq/SCM. The details of this measurement and the precise sensitivity analysis is reported in a separate publication (\cite{Aalseth2013}).

\subsection{GEANT simulations on cosmic ray backgrounds}
The above results from the surface measurement (\cite{Aalseth2011}) and the shallow underground sensitivity estimate (\cite{Aalseth2013}) lead to a desire to understand the impact cosmic rays have on the background in the proportional counters in the $^{37}$Ar measurement region. The relevant characteristics are the counter's length, radius, operating pressure, and the orientation of the cylindrical axis with respect to the vertical direction. For the moment, the orientation is fixed so the proportional counter axis is parallel the Earth's surface (i.e., the tube is horizontal) as this is how the initial above ground measurement was performed (\cite{Aalseth2011}). A simple GEANT model of an array of ULBPCs was prepared and cosmic-ray neutrons, protons, and muons were passed through the counter array. The Cosmic-ray Shower Library (CRY 1.5) (\cite{Hagmann2007} and \cite{Hagmann2008}) was used as input to GEANT to generate the cosmic ray neutrons, protons, and muons with approximately the correct relative proportion and angular distribution of incoming direction.


\begin{figure}
\begin{center}
\includegraphics[width=\columnwidth]{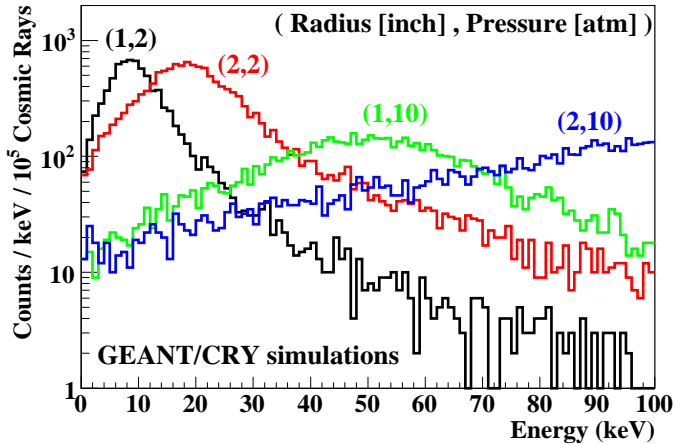}
\caption{\label{fig:CRYsim} Proportional counter optimization study. Cosmic rays generated from the CRY package are simulated in GEANT. The geometry and pressure of the proportional counter is varied. Shown here is the impact of radius and pressure on the totally energy deposited in the proportional counter gas. For an $^{37}$Ar measurement the sensitive energy region is 2--5~keV, thus designs that \emph{increase} the total energy deposited by cosmic rays assist in \emph{reducing} the background in the $^{37}$Ar measurement region. Note: 1 inch = 25.4 mm and 1 atm = 101.3 kpa.}
\end{center}
\end{figure}

For each of the simulations preformed, the integral number of counts in the energy range from 2 to 5 keV was determined and used as representative of the \emph{relative} importance of a particular proportional counter's length, radius, and operating pressure. In this way it is possible to ascertain the preferred proportional counter design for minimizing the cosmic ray background in an above ground system, in this example targeting measurement of the 2.8~keV signature of $^{37}$Ar decay. The results are presented in Tables \ref{tab:Length}, \ref{tab:Pressure}, and \ref{tab:Radius}. Figure \ref{fig:CRYsim} presents a sub-set of the simulations that illustrate the impact of counter radius and pressure on the measured background energy spectrum. The broad peak seen in each energy spectrum is essentially due to the amount of energy deposited along the ionizing particle's path length through the counter (i.e., ``minimum ionizing particle'' $dE/dx$). Thus, larger radius and higher pressure tubes show a broader spectral peak pushed to higher energy.

Putting aside other aspects of making a feasible, operational proportional counter, the summary conclusion from these simulation results is that the ideal counter is short in length, large in radius, and high in pressure. Note that large radius and high pressure creates a proportional counter configuration requiring very high potential bias for operation (i.e., high V/cm in the gas). In comparing the relative effect of pressure and radius, pressure has a greater impact. Thus a tube construction design driven by ability to achieve a desired high pressure, effectively selects the proportional counter radius by being required to keep the high voltage bias below 5000~V, a typical commercial high voltage bias supply module. Having fixed the operating pressure and radius to obtain proportional charge gain with a reasonable high voltage bias power supply, the final requirement is obtaining the desired sample-size target by selecting the proportional counter length.

{
\renewcommand{\arraystretch}{1.25}
\begin{table}
\begin{center}

\caption{\label{tab:Length} \textit{Length series.} Integral counts for the energy range 2--5~keV per $10^{5}$ simulated cosmic rays. Pressure = 2~atm and radius = 1~inch for all simulations in this table. Note the counts do not simply double as the length doubles do to an ''edge effect'' in the simulation since the cosmic rays were generated from a 1000 mm by 1000 mm surface centered above the proportional counter.}
\vspace{0.5em}
\begin{tabular}{lccccc}
\hline
Length (mm) & 100 & 200 & 300 & 400 & 500 \\
Counts (2--5 keV) & 596 & 1017 & 1334 & 1622 & 1776 \\
\hline
\end{tabular}

\vspace{0.5em}

\caption{\label{tab:Pressure}  \textit{Pressure series.} Integral counts for the energy range 2--5~keV per $10^{5}$ simulated cosmic rays. Length = 200~mm and radius = 1~inch for all simulations in this table.}
\vspace{0.5em}
\begin{tabular}{lcccc}
\hline
Pressure (atm) & 2 & 5 & 10 & 15  \\
Counts (2--5 keV) & 1017 & 167 & 59 & 37 \\
\hline
\end{tabular}

\vspace{0.5em}

\caption{\label{tab:Radius} \textit{Radius series.} Integral counts for the energy range 2--5~keV per $10^{5}$ simulated cosmic rays. Length = 200~mm for all simulations in this table.}
\vspace{0.5em}
\begin{tabular}{lccccc}
\hline
Radius (mm) \phantom{$\biggl[$} & 9.79 & 16.04 & 22.29 & 28.54 & 34.79 \\
\hline
\multicolumn{6}{l}{\emph{Pressure = 2~atm}} \\
Counts (2--5 keV) & 2134 & 1017 & 661 & 498 & 424 \\
\hline
\multicolumn{6}{l}{\emph{Pressure = 10~atm}} \\
Counts (2--5 keV) & 81 & 59 & 66 & 73 & 77 \\
\hline
\end{tabular}

\end{center}
\end{table}
}

\subsection{External gamma-ray backgrounds}
\label{subsec:ExternalGammas}
\begin{figure}
\begin{center}
\includegraphics[width=1.00\columnwidth]{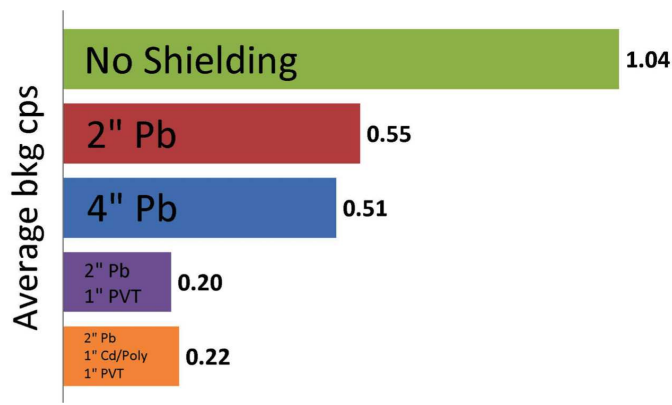}
\caption{\label{fig:AboveGroundOFHC} Results from a shielding study using an above ground OFHC copper proportional counter operated at 3 atm of P10 counting gas.}
\end{center}
\end{figure}

Above ground proportional counter measurements using a OFHC copper proportional copper were performed (\cite{MaceLRT2013}) to determine the affects of lead shielding on non-cosmic ray backgrounds. The shield used in this study was composed of 2$^{\prime\prime}$ thick lead in a cave of dimensions 8$^{\prime\prime}$$\times$8$^{\prime\prime}$$\times$16$^{\prime\prime}$  (W$\times$H$\times$L). Above and below the cave were 1$^{\prime\prime}$ thick sheets of cadmium (neutron absorber) and 1$^{\prime\prime}$ thick PVT scintillator panels (cosmic-ray veto). This shield design was then rebuilt increasing the lead shielding thickness to 4$^{\prime\prime}$. The OFHC copper proportional counter was also operated outside the shield in a typical laboratory bench-top environment. Figure \ref{fig:AboveGroundOFHC} shows the results of these comparative shielding studies. The average background rate was calculated over an energy range or 3-400 keV.

These results are consistent with the results previously discussed from an above-ground, shielded ULBPC (Sec.  \ref{subsubsec:SurfaceBackground}) (\cite{Aalseth2011}). The results presented in Fig. \ref{fig:AboveGroundOFHC} clearly demonstrate the need for lead shielding, though thicknesses of greater than 2$^{\prime\prime}$ do not appear to provide a large additional improvement in background. It seems reasonable to conclude from these results 2$^{\prime\prime}$ of lead shielding is more than adequate to shield against external ubiquitous primordial gamma-ray emitting radioactivity sources, though perhaps less lead could be employed (perhaps 1$^{\prime\prime}$) if the system's weight became a design consideration.

\subsection{Proportional counter construction material backgrounds}

To compare the backgrounds rates in an ULBPC\footnote{Specifically, PNNL counter number 2U03.} and an OFHC copper proportional counter\footnote{Specifically, PNNL counter number OFHC2.}, a test was performed in the PNNL shallow underground laboratory using the apparatus described in \cite{Aalseth2013} composed of cosmic ray and lead shielding arrangement. Spectroscopic histograms of the data from each of the two counters (both operated at $\sim$1 atm) show a generic shape indicative of leakage of un-vetoed cosmic ray induced backgrounds, both before and after the application of a cut from the cosmic ray veto system. The residual rates in each counter after application of the cosmic ray veto system were 187.3~$\pm$~1.8 cpd and 335.8~$\pm$~3.1 cpd in the ULPBC and OFHC copper counters, respectively. Although it would not be considered a ``conservative'' assumption, one could presume the ULBPC has essentially zero background counts from the copper material and there-by conclude the difference in the two rates is representative of the absolute background rate of a proportional counter constructed from OFHC copper at 148.5~$\pm$~3.6 cpd. Alternatively, the conservative assumption is of course to assume \emph{all} the residual background events are due to internal radioactive backgrounds in the construction materials, that is, a background rate of 335.8~$\pm$~3.1 cpd. These ranges of internal backgrounds from the OFHC copper proportional counter construction materials figure in estimating the lowest minimum detectable concentration (MDC) that can be achieve from a OFHC copper proportional counter operated in a surface laboratory.

\section{Absolute gas counting methods}
\label{sec:AbsoluteCounting}

In the case of a single proportional counter instrument, determining the absolute activity concentration of an unknown radioactive gas sample requires a calibration of the counter's efficiency. This efficiency is often driven by the details of the proportional counter design at the ends of the tube where electric fields lines may not produce collectible charge transport with the same proportional gain characteristics as the central portions of the tube. It is possible to use two (or more) separate but ``identical'' proportional counters to effectively remove any uncertainty related to the ends of the proportional tubes (\cite{Unterweger2007}). In summary explanation, assuming the end effects of the compared proportional counters are the same (hence the term ``identical''), subtraction of the shorter counter's results from the longer counter's results effectively creates a virtual proportional counter volume equal to the difference in the two counters' volumes but absent of any end effects.

\subsection{Method of variable length proportional counters}

The method of variable length proportional counters was explored previously using both OFHC copper proportional counters and ULBPC copper tubes (\cite{Williams2013}). The focus of that work was to develop the capability to directly prepare absolute activity concentration gas standards in the range 0.05-10~Bq/cm$^{3}$. In this report we consider the potential use of the same method of variable length proportional counters as a component of an quantitative assessment of sample gases. We perform this analysis with the same context of measurements of naturally occurring levels of $^{37}$Ar (\cite{Riedmann2011}). The reported concentration levels for soil gas are in the range of 1-100~Bq/SCM (SCM $\equiv$ standard cubic meter) of whole air. When 100\% of the argon is extracted from whole air, this results in a argon gas sample having an activity concentration in the range of approximately 1-100~$\mu$Bq/cm$^{3}$ -- dramatically lower than considered in our prior work (\cite{Williams2013}). Furthermore, we take into consideration the prior described background rates of the above ground and below ground proportional counters described in Sec. \ref{sec:ProportionalCounterBackgrounds}. Recall the Figures \ref{fig:ULBPCCosmicRayBackground} and \ref{fig:ULBPCCosmicRayBackgroundShallowLab} report background rates of 42~mHz (over ~0-20~keV) and 0.6~mHz (over 1-15~keV) for the 100~cm$^{3}$ volume proportional counters. A rough estimate of the background rate \emph{per keV} is then approximately 2~mHz/keV above ground to 0.04~mHz/keV for the shallow underground system. For the sake of providing a concrete sensitivity analysis, we presume we will be able to achieve a background rate in an above ground, well shielded proportional counter array of 0.5~mHz/keV/counter. This value is approximately a order of magnitude higher than achievable in a shallow underground laboratory (See Sec. \ref{subsubsec:ShallowBackground}), but a factor of 4 improvement in the background from the prior above ground system results (See Sec. \ref{subsubsec:SurfaceBackground}).

\subsection{Sensitivity and uncertainty analysis}
As done previously (\cite{Williams2013}), using the approach outlined in the Guide to the Expression of Uncertainty (GUM) (\cite{ISO1995}), an uncertainty model was constructed for sensitivity and uncertainty analysis using the GUM Workbench software tool (\cite{Metrodata-GmbH2012}). The model represents the 1$^{\prime\prime}$-diameter ULBPC tubes described in Sec. \ref{subsubsec:SurfaceBackground} (\cite{Aalseth2009}). Two equations are employed, the first for two different length proportional counters ($V_{\mathrm{S}}$ = 100~cm$^{3}$ and $V_{\mathrm{L}}$ = 250~cm$^{3}$ volumes) and the second for the larger ($V_{\mathrm{L}}$ = 250~cm$^{3}$ volume) tube alone. In this analysis, it is envisioned both proportional counters are loaded with an argon gas sample and 10\% methane by volume to make P-10, thus giving a mol fraction of $f_{\mathrm{mol}}$ = 0.90. Data is then collected for 24 hours (86400 seconds). Equation 1 is used to analyze the results from the two tubes jointly while Equation 2 uses the same data but only considers the results from the longer of the two tubes.
\begin{equation}\label{eq:one}
A_{\mathrm{L}-\mathrm{S}} = \frac{\left(1+k_{\mathrm{t}}+k_{\mathrm{w}}\right)\left(\frac{1}{f_{\mathrm{BR}}}\right)\left[\left(\frac{S_{\mathrm{L}}}{t}-\frac{B_{\mathrm{L}}}{t}\right)-\left(\frac{S_{\mathrm{S}}}{t}-\frac{B_{\mathrm{S}}}{t}\right)\right]}{T_{\mathrm{STP}}\left(\frac{P}{P_{\mathrm{STP}}}\right)f_{\mathrm{mol}}\left(\frac{1}{Z}\right)\left(\frac{V_{\mathrm{L}}}{T}-\frac{V_{\mathrm{S}}}{T}\right)}
\end{equation}
\begin{equation}\label{eq:two}
A_{\mathrm{L}} = \left(\frac{1}{\epsilon_{\mathrm{fid}}}\right)\frac{\left(1+k_{\mathrm{t}}+k_{\mathrm{w}}\right)\left(\frac{1}{f_{\mathrm{BR}}}\right)\left[ \left(\frac{S_{\mathrm{L}}}{t}-\frac{B_{\mathrm{L}}}{t}\right)\right]}{T_{\mathrm{STP}}\left(\frac{P}{P_{\mathrm{STP}}}\right)f_{\mathrm{mol}}\left(\frac{1}{Z}\right)\left(\frac{V_{\mathrm{L}}}{T}\right)}
\end{equation}
In these equations, $k_{\mathrm{t}}$ is the ratio of those decays \emph{not} measured to those decays that are measured. In this particular example that amounts to an assessment of the fraction of events contributing to the signal region of interest (ROI) for analysis. In a typical analysis, a Gaussian distribution is fit to the energy peak located at 2.8~keV. The central 82.6\% of the peak distribution is used as the ROI thus giving a $k_{\mathrm{t}}$ value equal to (100 - 82.6)/82.6 = 0.21. A similar quantity, $k_{\mathrm{w}}$ is the ratio of those decay events that range out in the walls of the proportional coutner to those decay events that create a signal. In this particular case, it is assumed that nearly all (99\%) of the $^{37}$Ar decays produce event signals as the energy deposition is low energy and nearly point-like in the proportional count gas described here. Thus the $k_{\mathrm{w}}$ is  (100 - 99)/99 = 0.01. The uncertainty for these values is estimated relative to our expectations for how well we would be able to determine these values from our current measurement experiences. The percentage of the fiducial volume of the longer proportional counter that is sensitive to $^{37}$Ar decay is $\epsilon_{\mathrm{fid}}$. The branching fraction of the $^{37}$Ar decay to the K-shell is $f_{\mathrm{BR}}$ and is considered a known constant for the purposes here. The count duration, $t$, is given in seconds. The proportional counters are assumed to operate at a pressure $P$ = 709.275~kPa and at a temperature $T_{\mathrm{L}}$ = $T_{\mathrm{S}}$ = 300~K. The compressibility of P-10 gas is $Z$ = 0.998 and is considered a known constant. All of these fixed values are summarized in Table~\ref{tab:GUM}.

{
\renewcommand{\arraystretch}{1.25}
\begin{table}
\begin{center}
\caption{\label{tab:GUM} Fixed inputs for the GUM uncertainty model. The volume values $V_{\mathrm{L}}$ and $V_{\mathrm{S}}$ are correlated via determination from the same calibrated reference.}
\vspace{0.5em}
\begin{tabular}{cccc}
\hline
Quantity & Unit & Value & Uncertainty \\
\hline
$k_{\mathrm{t}}$      &  -  & 0.21 & 0.010 \\
$k_{\mathrm{w}}$     &  -  & 0.01 & 0.005 \\
$\epsilon_{\mathrm{fid}}$ & \% & 94.0 & 0.5 \\
$f_{\mathrm{mol}}$  & mol/mol & 0.90 & 0.01 \\
$t$                           &  s & 86400 &  1  \\
$P$                           & kPa & 709.275 & 0.2 \\
$V_{\mathrm{L}}$     & cm$^{3}$ &  250  &  0.625 \\
$V_{\mathrm{S}}$     & cm$^{3}$ &  100  &  0.250 \\
$T$                          &  K  & 300 &  0.1 \\
$f_{\mathrm{BR}}$    & \% & 90.2 &  constant  \\
$Z$                          &  -  & 0.998 &  constant  \\
$T_{\mathrm{STP}}$  &  K  & 273.15 &  constant  \\
$P_{\mathrm{STP}}$  &  kPa & 101.325 &  constant  \\
\hline
\end{tabular}
\end{center}
\end{table}
}

The number of counts measured in the signal ($S$) region of interest (ROI) and the number of counts in the background ($B$) determined from side band regions away from the tails of the Gaussian distribution fit are represented in the equations by $S_{\mathrm{L}}$, $B_{\mathrm{L}}$, $S_{\mathrm{S}}$, and $B_{\mathrm{S}}$ for the long (subscript $\mathrm{L}$ $\rightarrow$ 250~cm$^{3}$) and short (subscript $\mathrm{S}$ $\rightarrow$ 100~cm$^{3}$) proportional counters. The following comparative analysis, the signal counts are estimated for three difference measurement cases of $^{37}$Ar activity concentration in whole air: 1000~mBq/SCM, 100~mBq/SCM, and 10~mBq/SCM. As described above, the background rate is assumed to 0.5~mHz/keV over an 0.8 keV width of energy region that is equivalent to the width of the 82.6\% of the peak distribution is used as the signal ROI. The estimated counts in each energy region for each proportional counter are tabulated in Table~\ref{tab:SignalBackgroundValues} for the three $^{37}$Ar activity concentration cases.

{
\renewcommand{\arraystretch}{1.25}
\begin{table}
\begin{center}
\caption{\label{tab:SignalBackgroundValues} Estimated signal region ($S$) and background region ($B$) counts for example 24-hour measurements described in the text for long ($L$) and short ($S$) proportional counters filled with argon extracted from whole air. Activity concentrations are given in mBq per standard cubic meter (SCM) of whole air.}
\vspace{0.5em}
\begin{tabular}{c|c|c|c}
\hline
Count Region  & \multicolumn{3}{c}{Activity (mBq/SCM)} \\
                      &   10  &  100  &   1000 \\
\hline
$S_{\mathrm{L}}$ & 180 & 1008 & 9288 \\
$B_{\mathrm{L}}$ &   88 &   88   & 88 \\
\hline
$S_{\mathrm{S}}$ & 68  & 365 & 3335 \\
$B_{\mathrm{S}}$ & 35  & 35  & 35 \\
\hline
\end{tabular}
\end{center}
\end{table}
}

The GUM Workbench software tool (\cite{Metrodata-GmbH2012}) output provides an \emph{index} of the uncertainty contribution of a given input value and input value uncertainty. The higher the index percentage for a given input value and input value uncertainty, the more that input and its uncertainty \emph{drives} the overall calculated uncertainty of the equation modeled. From Table~\ref{tab:SignalBackgroundValues} one sees the ``signal-to-noise'' ratio ranges from $\sim$1 (for 10~mBq/SCM $^{37}$Ar in the short counter) to $\sim$100 (for 1000~mBq/SCM $^{37}$Ar in the long counter).

{
\renewcommand{\arraystretch}{1.25}
\begin{table}
\begin{center}
\caption{\label{tab:Index} The GUM Workbench index of contribution to uncertainty reported for various input parameters modeled in equations \ref{eq:one} and \ref{eq:two}. The index of contribution to uncertainty is the relative contribution the input parameter and input parameter's uncertainty contribution to the \emph{overall} uncertainty reported on $A_{\mathrm{L}-\mathrm{S}}$ or $A_{\mathrm{L}}$. The index of contribution to uncertainty is reported on a scale from 0.0-1.0 where larger values indicate a \emph{greater} contribution to the uncertainty of the modeled equation. The quantities $A_{\mathrm{L}-\mathrm{S}}$ and $A_{\mathrm{L}}$ are reported in $\mu$Bq/cm$^{3}$. For conversion purposes, 10~mBq of $^{37}$Ar per standard cubic meter (SCM) of whole air is equivalent to 1~$\mu$Bq/cm$^{3}$.}
\vspace{0.5em}
\begin{tabular}{c|ccc|ccc}
\hline
\multicolumn{7}{c}{Relative uncertainty on reported activity concentration (\%)} \\
\hline
                     & \multicolumn{3}{c|}{Length Compensated} & \multicolumn{3}{c}{Single Long Counter} \\
                     & \multicolumn{3}{c|}{Activity (mBq/SCM)} & \multicolumn{3}{c}{Activity (mBq/SCM)}\\
                     &   10  &  100  &   1000  &   10  &  100  &   1000 \\
\hline
$A_{\mathrm{L}-\mathrm{S}}$ & 65 & 13 & 4.4 &  &  &  \\
 $A_{\mathrm{L}}$                   &  &  &  & 36 & 7.6 & 3.2 \\
\hline
\multicolumn{7}{c}{} \\
\hline
\multicolumn{7}{c}{Index of contribution to concentration uncertainty} \\
\hline
$S_{\mathrm{L}}$ & 0.485 & 0.656 & 0.551 & 0.669 & 0.828 & 0.433 \\
$B_{\mathrm{L}}$ & 0.237 & 0.057 & 0.005 & 0.327 & 0.072 & 0.004 \\
\hline
$S_{\mathrm{S}}$ & 0.183 & 0.237 & 0.198 & - & - & - \\
$B_{\mathrm{S}}$ & 0.094 & 0.023 & 0.002 & - & - & - \\
\hline
$k_{\mathrm{t}}$               & 0.000 & 0.015 & 0.139 & 0.002 & 0.047 & 0.265 \\
$\epsilon_{\mathrm{fid}}$ & - & - & - & 0.000 & 0.017 & 0.099 \\
$f_{\mathrm{mol}}$           & 0.000 & 0.009 & 0.085 & 0.001 & 0.029 & 0.162 \\
\hline
\end{tabular}
\end{center}
\end{table}
}

The results from Table~\ref{tab:Index} are informative because over the range of relevant measurement activity concentrations, the contribution to uncertainty in the determination from a given measurement shifts from a statistics limited situation (for low activity concentration) to systematics limited situation (for higher activity concentrations). Perhaps surprisingly the analysis selection of the energy range over which the $^{37}$Ar decay events are counted (the analysis ROI) becomes an important consideration in the higher activity concentration measurements. Stated another way, the knowledge of the Gaussian width of the $^{37}$Ar decay peak -- and the uncertainty associated with that knowledge -- impacts how well known the fraction of the peak covered by the analysis ROI. An interesting aside is, for this example case, one can consider the choice between using the length compensated method or a single well calibrated counter as a choice between having the measurement uncertainty driven by \emph{either} the statistics of the short counter in the length compensated method (i.e., $S_{\mathrm{S}}$ and $B_{\mathrm{L}}$) \emph{or} the systematic uncertainties associated with knowledge of the ``sample size'' (i.e., $k_{\mathrm{t}}$, $\epsilon_{\mathrm{fid}}$, and $f_{\mathrm{mol}}$).

\section{Proportional counter array design}
\label{sec:DesignConcept}

\begin{figure}
\begin{center}
\includegraphics[width=0.75\columnwidth]{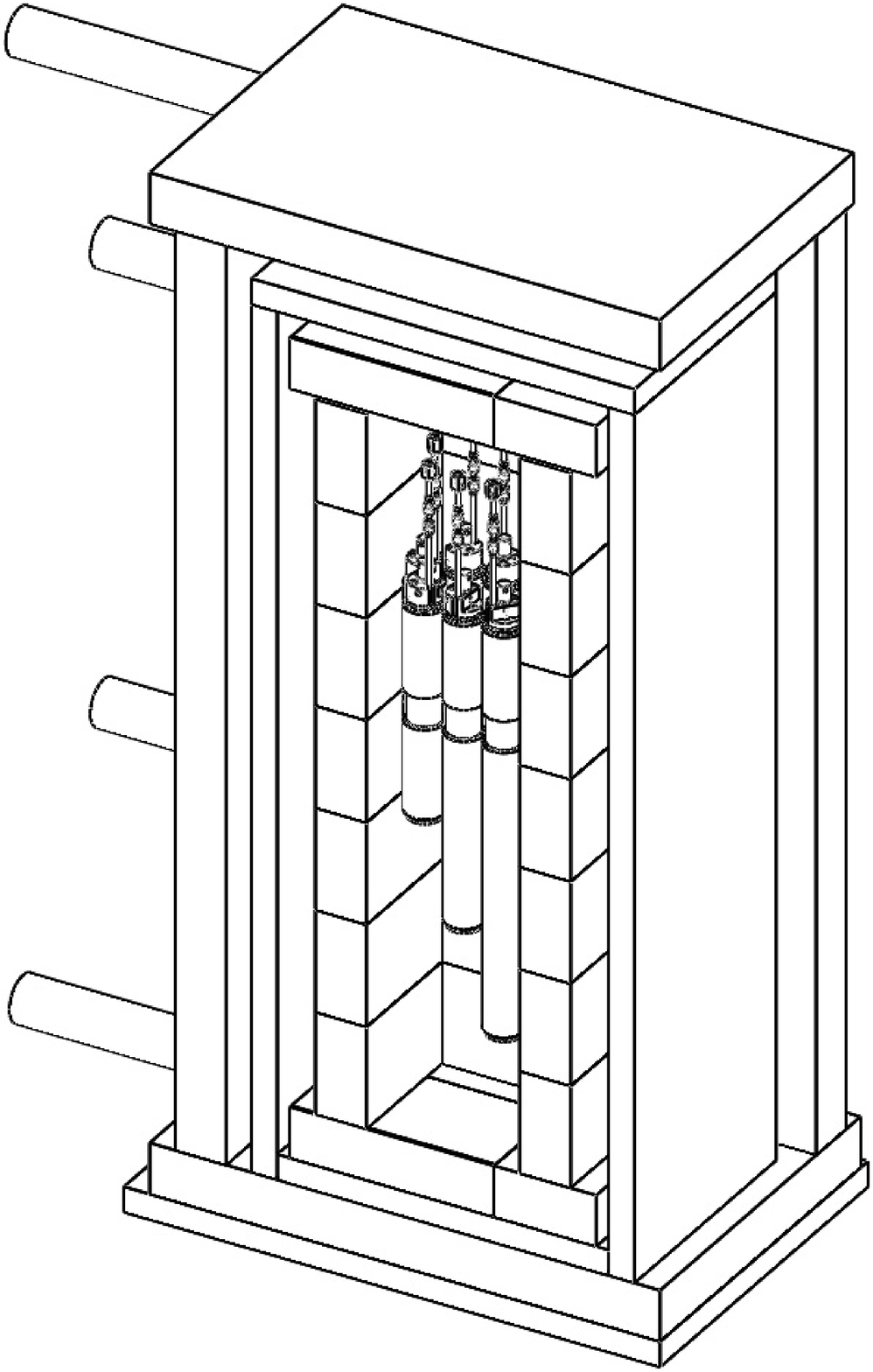}
\caption{\label{fig:Ar37Shield} Concept for a shield intended for low background gas proportional counters located in surface laboratories.}
\end{center}
\end{figure}

Figure \ref{fig:Ar37Shield} presents one conceptual realization of the cosmic ray background reduction and mitigation presented in this article. The individual proportional counters are oriented vertically as this will typically increase the path length through the gas, there-by pushing the energy measured from cosmic-ray products further away from the low energy $^{37}$Ar signature at 2.8~keV. A closely packed array of proportional counters increases the ability to veto events based upon coincidence triggers between the tubes\footnote{In the low-rate configurations studied here, the accidental coincidence rate from the $^{37}$Ar decays is inconsequential.}. The inner shield shown in Fig. \ref{fig:Ar37Shield} is composed of 2$^{\prime\prime}$-thick standard lead bricks to eliminate external gamma-rays (See Sec. \ref{subsec:ExternalGammas}). The lead shield is surrounded by 1$^{\prime\prime}$-thick borated polyethylene to provide some thermal neutron absorption. Finally, and most importantly, the entire shield is enclosed in a active plastic scintillator veto system. The scintillator panels are 2$^{\prime\prime}$-thick to ensure the ability to separate muons from gamma rays. This shield as design is conceptual and weights approximately 1400 pounds (635 kg). The choice of variable proportional counter length is schematic and depends on the final measurement goal, sample volume, and activity concentration.

\section{Gas preparation example}
\label{subsec:artestbench}

\begin{figure*}[ht]
\begin{center}
\includegraphics[width=2\columnwidth]{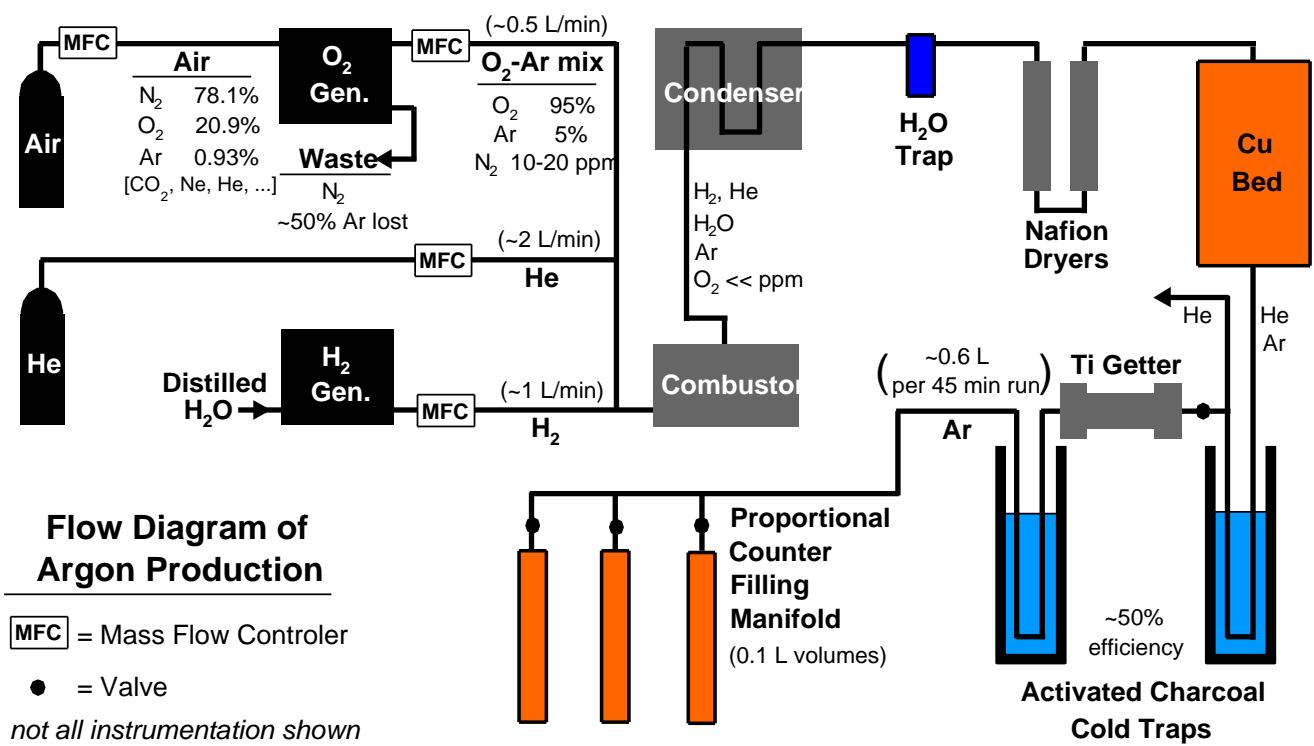} 
\caption{\label{fig:ArgonProduction} Flow diagram for the production of argon from an air sample.}
\end{center}
\end{figure*}

In this article we have considered the measurement of $^{37}$Ar as a concrete example to study shielding requirements for low-background gas proportional counters located in above ground (surface) laboratory settings where cosmic rays are expected to dominate the background of the system. As briefly mentioned earlier in this article, the choice of the proportional counter \textit{length} is the least background-dependent handle available to adjust the system's capacity relative to a given sample preparation methodology. To further discuss this sample-size aspect of system design considerations, we again reference $^{37}$Ar measurements in the context of determination of the naturally occurring levels of $^{37}$Ar in soils types and at depths within those soils (\cite{Riedmann2011}). We present a concept for processing a soil gas sample for $^{37}$Ar measurement as a specific example of a potential gas sample size to assist in understanding the capacity requirements of gas proportional counters as described in this article.

For the purposes of this article, we assume any dried soil-gas sample is represented in rough volumetric composition by normal dry air: N$_{2}$ (78\%), O$_{2}$ (21\%), Ar (0.9\%), CO$_{2}$ (0.04\%), and other minor constituents each at less than 0.01\%. The measurement goal is to determine the $^{37}$Ar concentration in $\sim$mBq/SCM (\cite{Riedmann2011}). Although a variety of gas mixtures are possible for gas proportional counters methods, a natural choice for an $^{37}$Ar measurement is P-10 (90\% argon and 10\% methane), where the 90\% argon fraction is supplied from the soil-gas sample.

As a general example, the ultra-low-background proportional counters (ULBPCs) hold $\sim$0.1~L of gas volume when filled at atmospheric pressure and up to $\sim$1~L of gas volume when filled at 10 times atmospheric pressure. Although the ULBPC typically operates at fill pressures between these two extremes, we first consider as a simple example the $\sim$1~L of gas volume, 10 atmosphere fill case. In this case, $\sim$90\% of the fill gas is argon or $\sim$0.9~L of argon at atmospheric pressure. To obtain 0.9~L of argon from dried air (e.g. the soil-gas sample), 100~L (or 0.1~SCM) of dried air is required as an initial sample collection, assuming 100\% collection efficiency when separating the argon from the desiccated air. A 100\% collection efficiency of argon gas from dry air is unrealistic. Determining the actual representative argon collection efficiency is of primary interest as well as understanding the processing time.

To this end, an argon production test bench was developed to quantify the aspects of argon production efficiency as well as processing duration. A schematic flow diagram of the system is presented in Figure \ref{fig:ArgonProduction}. The primary step in the separation of argon from dry air is a pressure swing absorption (PSA) oxygen (O$_{2}$) generator. The PSA O$_{2}$ generator eliminates the predominate nitrogen (N$_{2}$), producing an oxygen-argon mixture (O$_{2}$-Ar mix), but at the loss of approximately 50\% of the argon in the initial air sample. After the output of the oxygen (O$_{2}$) generator is collected, the O$_{2}$-Ar mix is fed into a carrier gas of helium (He) and then mixed with hydrogen (H$_{2}$) for converting the hydrogen and oxygen into water vapor (H$_{2}$O). The water is collected and the gas stream composed of hydrogen, helium, argon, and trace oxygen and nitrogen is dried and passed through a reactive copper bed. The final gas stream is composed of argon and helium, that are finally separated via activated charcoal traps where a nominal 50\% efficiency is assumed. In a typical process cycle, approximately 0.6~L of argon is produced from 150~L of initial air sample. This process cycle is composed of approximately 45~minutes of active processing. An entire cycle including regeneration time, requires approximately 2~hours. Thus, in this argon production test bench cycle approximately 0.6~L of argon sample is prepared every 2~hours. This sample size and production rate is on the scale relevant to the 24-hour count duration of $\sim$1~L volumes described in Sec. \ref{sec:AbsoluteCounting}.

\section{Conclusion}
\label{sec:Conclusion}

The value of ultra low background proportional counters (and low background radiation detection systems in general) is typically achieved when applied to measurements made in underground laboratories. In such underground systems, where cosmic rays have been eliminated as a major contributor to the background of the system, the uranium, thorium, and potassium content of the instrument components typically dominates. Conversely, developments of low background proportional counters have led to instruments of such low internal background, that if they were to be operated on the surface, the internal background contaminants effectively have zero contribution to the system's background. In such a case, the instrument and shielding design can focus exclusively on ways to mitigate the impact of cosmic rays products on the measurement. We have presented a conceptual thought process on how to best reduce and mitigate against cosmic ray background contributions in a surface laboratory setting for gaseous proportional counters. We analyzed the potential performance of such a system in the context of a simple radioactive gas, $^{37}$Ar, at concentration levels relevant to environmental measurements. We conclude it is potentially feasible to design and build an actively and passively shielded, proportional counter array that is sensitive to the range of naturally occurring $^{37}$Ar soil gas activity concentration levels in an above ground measurement system.

\section*{Acknowledgments}
\label{sec:Acknowledgments}

The research described in this paper was supported in part by the Ultra-Sensitive Nuclear Measurements (USNM) Initiative, a Laboratory Directed Research and Development Program at the Pacific Northwest National Laboratory, a multiprogram national laboratory operated by Battelle for the U.S. Department of Energy. Information Release PNNL-SA-106350.

\bibliographystyle{shieldopt-model2-names}
\bibliography{shieldopt-elsarticle-template-2-harv}







\end{document}